\documentclass[10pt]{article} 
\usepackage{amssymb,euscript,latexsym,amsmath} 
\usepackage{graphicx} 
\usepackage{enumerate} 
\usepackage{color} 
\usepackage{setspace} 
\usepackage{subfigure} 
\usepackage[authoryear]{natbib} 
\usepackage{hyperref}

\begin{document}

\title{\bf Coins falling in water} 

\author{Luke Heisinger, \   Paul K. Newton \ and Eva Kanso\footnote{Corresponding author: kanso@usc.edu}  \\[2ex]
 University of Southern California, Los Angeles, CA  90089 }

\maketitle

{\small
\singlespacing

\begin{abstract}   
When a coin falls in water, its trajectory is one of four types determined by its dimensionless moment of inertia $I^\ast$ and Reynolds number Re: (A) steady; (B) fluttering; (C) chaotic; or (D) tumbling. The dynamics induced by the interaction of the water with the surface of the coin, however, makes the exact landing site difficult to predict a priori. Here, we describe a carefully designed experiment in which a coin is dropped repeatedly in water to determine the probability density functions (pdf) associated with the landing positions for each of the four trajectory types, all of which are radially symmetric about the centre drop-line. In the case of the steady mode, the pdf is approximately Gaussian distributed with small variances, indicating that the coin is most likely to land at the center, right below the point  from which it is dropped. For the other falling modes, the center is {one of the least} likely landing sites. Indeed, the pdf's of the fluttering, chaotic and tumbling modes are characterized by a ``dip" around the center. \textcolor{black}{In the tumbling mode, the pdf is a ring configuration about the center-line 
whereas in the chaotic mode, the pdf is generally a broadband distribution spread out radially symmetrically about the center-line. For the steady and fluttering modes, the coin never flips, so the coin lands with the same side up as when it was dropped. The probability of heads or tails is close to 0.5 for the chaotic mode and, in the case of the tumbling mode, the probability of heads or tails {is} based on the height of the drop which determines whether the coin flips an even or odd number of times during descent.}
\end{abstract}
}

\singlespacing
\section{Introduction}

Predicting the landing site of an object descending under the influence of gravitational and aerodynamic forces is relevant to many branches of engineering and science. Examples range from estimating the touchdown locations of re-entry space vehicles (e.g.,~\cite{ZhJi2006}) to studying seed dispersal and its influence on plant ecology (e.g.,~\cite{HoSm1982,NaMu2000}).

The descent motion of freely falling objects is generally complex,  even for regularly shaped objects such as coins and cards that flutter (oscillate from side to side) or tumble (rotate and drift sideways). 
A qualitative description of the tumbling motion dates back to the work of~\cite{Maxwell1854}. More recently,  experimental studies of the tumbling motion of freely falling plates was conducted by~\cite{Dupleich1941} who measured  the angle of descent and the average tumbling frequency as functions of the plate loading and the ratio between plate length and width.~\cite{Willmarth1964} constructed a phase diagram using careful experiments for falling discs with steady descent, fluttering, and tumbling. They found, in the limit of small thickness-to-width ratio, that the final state depends only on the Reynolds number and the dimensionless moment of inertia.  A  qualitatively similar phase diagram but for falling plates was constructed by~\cite{Smith1971}.~\cite{Field1997} conducted further experiments on freely falling discs and reported chaotic motion that marks the transition from fluttering to tumbling. The phase diagram of~\cite{Field1997}  is reproduced here in Figure~\ref{fig:fallingregimes}(left).  The transition from fluttering to tumbling was quantified in~\cite{BeEiMo1998} in a quasi two-dimensional experiment.


{\cite{PeWa2004} and Anderson, Pesavento \& Wang (2005a, 2005b) studied the motion of falling plates using a combination of  experimental measurements, direct numerical simulations of the two-dimensional Navier-Stokes equation, and an ordinary differential equation (ODE) model based on approximations of the fluid forces and torques in terms of lift and drag coefficients. Their interest in this problem stemmed from its relevance to passive flight, as opposed to flight by flapping wings, as well as to its importance as a testbed for the unsteady, hydrodynamic force models. Their system depended on three dimensionless parameters: the thickness-to-width ratio, the dimensionless moment of inertia, and the Reynolds number. They observed a transition from periodic fluttering to periodic tumbling as the thickness-to-width ratio increases (thereby also increasing the dimensionless moment of inertia). In the transition region, the period of oscillation diverges, and the ODE model exhibits a heteroclinic bifurcation that marks this transition and leads to a logarithmic divergence of the period of oscillation at the bifurcation point. Investigations of the fluttering and tumbling dynamics based on the numerical solution of the two-dimensional Navier-Stokes equations were also carried out by~\cite{MiSeUd2004}.}

\begin{figure}
\begin{center}
	\begin{subfigure}[Disc orientation]
		{\includegraphics[width=0.25\textwidth]{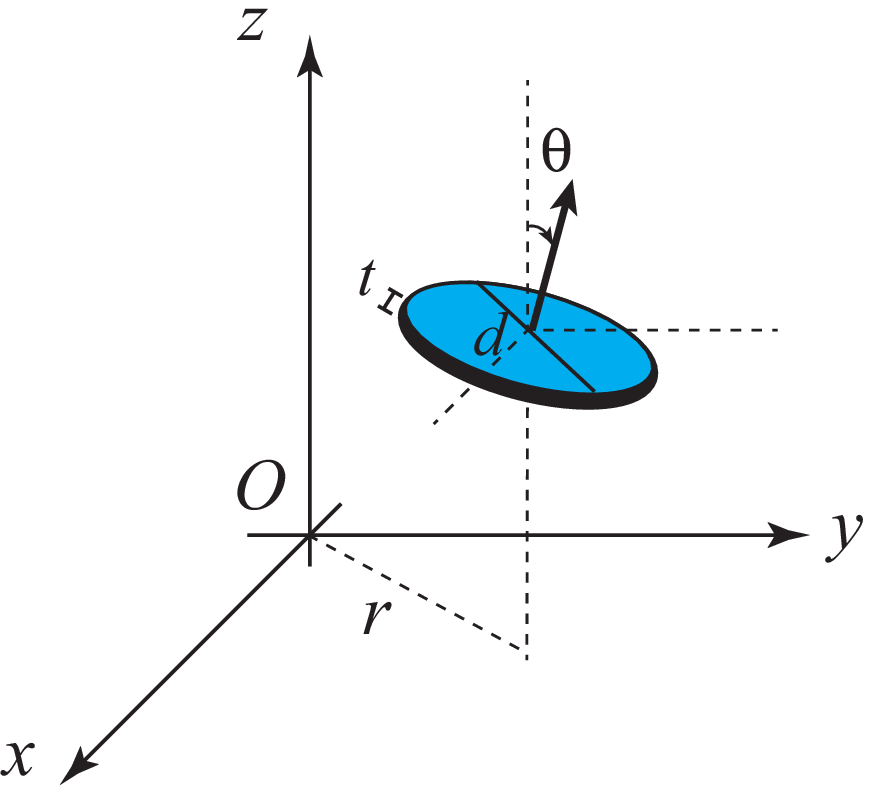}}	\hspace{0.25in}
	\end{subfigure}
	\begin{subfigure}[Water tank]
		{\includegraphics[width=0.25\textwidth]{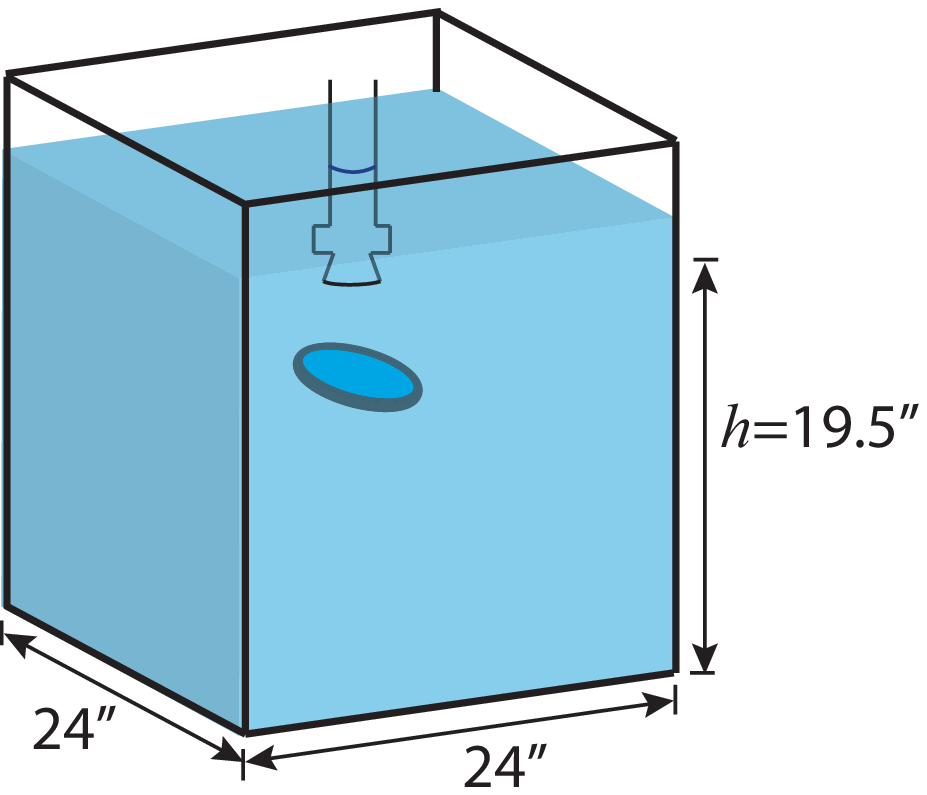}}	\hspace{0.25in}
	\end{subfigure}
	\begin{subfigure}[Camera set-up]
		{\includegraphics[width=0.25\textwidth]{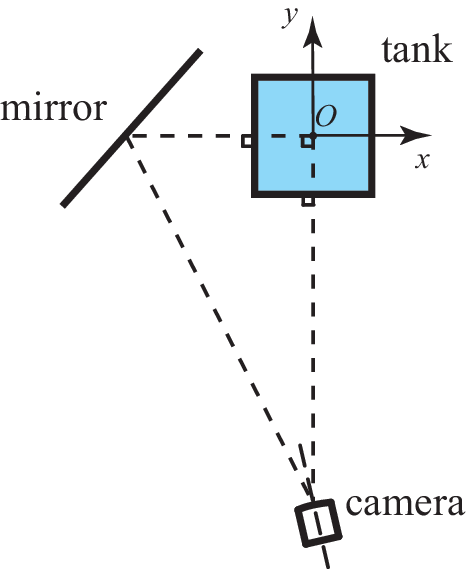}}	
	\end{subfigure}
	{\caption{schematic of the coin drop and experimental set-up}  } 
	\label{fig:schematic}
\end{center}
\end{figure}

{\cite{jonesshelley05} proposed a falling card model based on inviscid  flow theory and the unsteady Kutta condition with a vortex-sheet model of the wake, whereas~\cite{michel09} used an unsteady point-vortex model of the wake. These studies provided insight into the role of the vortical wake in the destabilization process that marks the transition away from the fluttering regime, albeit in two-dimensional flows.}

{\cite{Mahadevan1996} and \cite{Mahadevan1999} observed a scaling law for the dependence of the rotational speed on the width of a tumbling card. 
The effect of flexibility on the tumbling motion of a rectangular plate was analyzed by~\cite{TaBu2010} who found that flexibility leads to diminished flight characteristics, namely, an increased descent rate and a decreased flight range. Flexible sheets falling in an inviscid fluid were also considered by~\cite{Alben2010}  using techniques similar to those of~\cite{jones} and \cite{jonesshelley05}.  Alben reported statistical results linking the falling behaviors to a dimensionless rigidity parameter and the ratio of sheet to fluid densities.}

In this paper, we describe an experiment of falling coins in water with fixed initial conditions (up to the limits of the experimental set-up), shown in Figure~\ref{fig:schematic}. 
Depending on parameter values, the falling motion can
be described by four canonical trajectory types: (A) steady; (B) fluttering; (C) tumbling; or (D) chaotic. 
The goal of this experiment is to map out the probability density function (pdf) of landing sites associated with these four canonical trajectory types,
as well as the probability of landing the same side up (heads).

There are five material parameters of the coin/fluid system that determine the type of falling motion:  the disc diameter $d$, thickness $t$ and density $\rho$, as well as the fluid density $\rho_f$ and kinematic viscosity $\nu$. In addition, the height $h$ from which the disc is dropped plays a role in determining the landing site and the probability of landing heads or tail. The mean vertical  velocity $U$ of the disc is not a material parameter, but is rather an outcome of the falling dynamics and is not known a priori. {Here, we compute $U=h/T$, where $T$ is the total travel time, that is to say, the total time the disc takes to reach the bottom of the tank from its initial height $h$}.
From these seven physical variables, four independent dimensionless ratios may be formed.  The first quantity is the dimensionless moment of inertia {$I^\ast= I_{\mathrm disc}/ \rho_f \, d^5 = \pi \rho t/64 \rho_f d$}.  A second dimensionless quantity is the Reynolds number ${\rm Re} = U d/ \nu$.  The last two quantities are $t/d$ and $h/d$. For the discs we consider here $t/d$ is small {($t/d$ is of order $0.01$)} and does not play a significant role in the disc dynamics.  The dimensionless height $h/d$ is set to a constant, { $h/d=19.5\pm 0.125$}, determined by the size of the experimental apparatus. {Generally,   if  $h/d$ is large enough to allow the long-term descent dynamics to develop (as opposed to only the transient dynamics), its value does not affect the type of descent motion.  In the reported experiments,  we verified by experimenting with various heights that, for  the chosen disc parameters, the choice of $h/d = 19.5$ is sufficient for the long-term dynamics to develop in the fluttering and tumbling modes.} To this end, the type of motion is effectively determined by two parameters $I^\ast$ and ${\rm Re}$. This parameter space was explored  experimentally by~\cite{Field1997}. The authors
mapped out the four types of falling motion onto this parameter space. Their data suggest that the 
transition from fluttering to tumbling motion is marked by a region of chaotic behavior. 
The resulting `phase diagram' is reproduced here in Figure~\ref{fig:fallingregimes}(a). We added the dashed line separating the steady and tumbling motion because {for discs with large inertia, the gravitational acceleration dominates over the hydrodynamic forces} and one expects steady falling. {Note that the exact location of the dashed line remains to be validated by further experiments.}

\begin{figure}
\begin{center}
		{\includegraphics[width=0.9\textwidth]{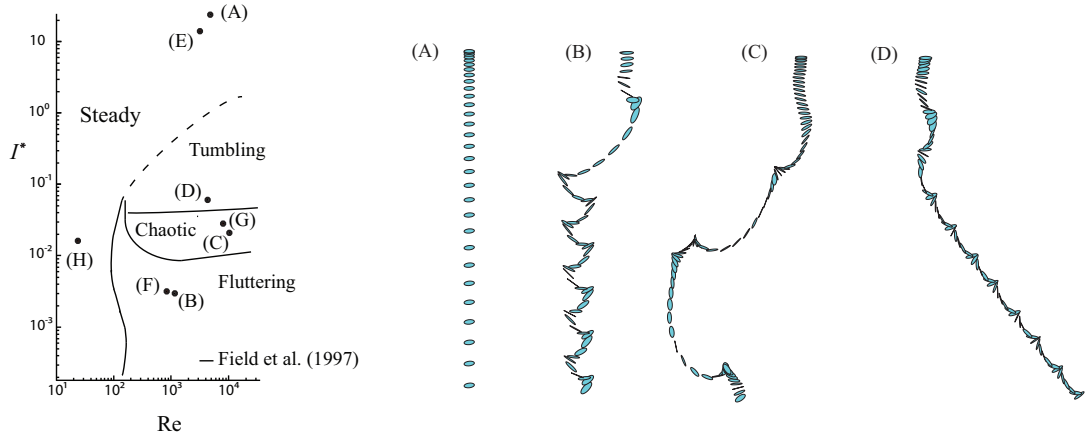}} \hspace{0.75in}
	\caption{Falling regimes: (left) Parameter space adapted from~\cite{Field1997}, spanned by the dimensionless moment of inertia and Reynolds number. The
dashed line indicates a parameter boundary anticipated by the authors that remains to be
validated by further experiments.
	 {{\rm (A)}  $d=1''$, $t=0.057''$ steel  in air. {\rm (B)} $d=1''$, $t=0.053''$  acrylic  in water. {\rm (C)}  $d=1''$, $t=0.057''$ steel in water. {\rm (D)} $d=0.5''$, $t=0.060''$ steel   in water.  The machining tolerance  is $\pm 0.002''$.} (right) Sample trajectories from each falling regime reconstructed {from our experimental data. Note the similarity with the trajectories reported in~\cite{Field1997}}.}
	\label{fig:fallingregimes}
\end{center}
\end{figure}

We consider four distinct points (A) through (D) of the parameter space, each representing a different type of falling motion as shown in Figure~\ref{fig:fallingregimes}. 
These points are obtained using steel and acrylic discs of various diameters in air and water. 
In all experiments, the coin is released from the same height below the fluid surface using the same release mechanism, {as described in Section~\ref{sec:methods}}. 
For each point (A) through (D), we drop the coin approximately 500 times repeatedly.
Interestingly, for the same parameter values and same initial conditions, up to a small uncertainty introduced by the drop mechanism, 
distinct drops of the coin result in distinct falling trajectories and landing positions. {In Section~\ref{sec:results}, we quantify the uncertainty introduced by the drop mechanism,} and
we report and explain the pdf's associated with the landing positions of these falling coins as well as their probability of landing heads or tails. 
{Our results show that the problem of a coin falling in water belongs to a class of deterministic problems -- that obey the laws of mechanics -- which generate outcomes that are best treated probabilistically. We conclude in Section~\ref{sec:conc} by discussing the link between the probabilistic outcome of the coin and its underlying mechanical motion, and on the role of the fluid medium as a ``randomization device."}

\section{Methods}
\label{sec:methods}

The experimental design is shown in Figure \ref{fig:schematic}, with Figure \ref{fig:schematic}(a) depicting the disc orientation \textcolor{black}{$\theta$, which is the angle between the disc's normal direction and the vertical direction}; Figure \ref{fig:schematic}(b) the water tank used in the experiment, and Figure \ref{fig:schematic}(c) the digital camera set-up.
In water, we dropped the disc from about {$1.2''$} below the top surface to avoid effects associated with {the water surface} and entry across the air-water interface, and we released the coin with zero initial velocity, heads-up.  A 24-inch cubic tank was used, and the drop position was centered such that the disc was always free to fall unconstrained and unaffected by edge effects from the walls. {We used steel and acrylic discs of respective densities $0.28$ lb/in$^3$ and $0.04$ lb/in$^3$, diameters $0.5''$ and $1.0''$, and  thickness values between $0.05''$ and $0.06''$ (see Figure~\ref{fig:fallingregimes})}.  For each disc diameter, a suction cup at least the diameter, or smaller, was used to suspend the disc below the surface of the water. 
The suction cup was attached rigidly to a tank lid structure, and a tube connected the submerged cup end to a rubber bulb. This bulb was used to create a vacuum while affixing the various discs to the suction cup, and enabled the manual release of the disc. 
A thin metal mesh was placed along the bottom to ensure that  the disc did not slide when it struck the bottom. 
The landing position was recorded by a digital camera mounted on the ground, facing directly upwards, with the focal plane coinciding directly with the bottom of the tank.  The camera was triggered remotely after each coin release to record its landing position. 
	
We also collected data on the three-dimensional position and orientation of the discs as they descended. A single camera was used in conjunction with a mirror to capture two views of the falling disc . The camera was not placed below the tank as before, but on a tripod level with the midpoint of the tank to minimize lens curvature distortion (see Figure 1(c) for camera set-up). Careful attention was paid to the mirror/camera/tank alignment to make the two views orthogonal to one another for digital processing and trajectory reconstruction.

\begin{figure}
\begin{center}
{\includegraphics[width=1\textwidth]{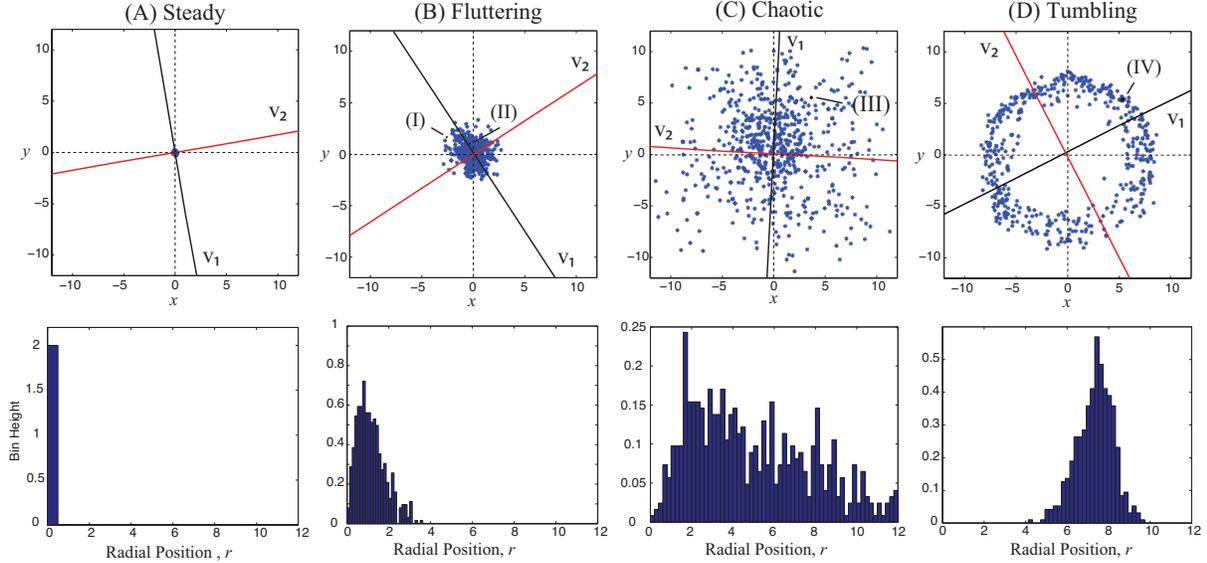}}
\caption{\footnotesize (top row) Distribution of landing sites for (A) steady, (B) fluttering, (C) chaotic and (D) tumbling motions. Each set of data correspond to one point in the parameter space shown in Figure~\ref{fig:fallingregimes}(left). (bottom row) histograms of the radial distribution of the landing positions. All length units are in inches.
\label{fig:distribution}}
\end{center}
\end{figure}

\begin{table}
\begin{center}
    \caption{Distribution parameters: $\sigma_{1,2}$ are the standard devations along the major
    and minor axes $V_1$ and $V_2$ of Figure~\ref{fig:distribution}, $\beta_{1,2}$ and $\gamma_{1,2}$ are
    the normalized skewness and kurtosis, respectively.} 
    \begin{tabular}{|  c |  c  c  c c  | }
    \hline
 &   \hspace{0.1in}  $N_{drops}$    \hspace{0.1in}        
 & \hspace{0.1in}  $P_{head}$   \hspace{0.1in} 
 &       \hspace{0.1in}  \hspace{0.1in} 
 &   \hspace{0.1in}  distribution parameters     \hspace{0.1in}                   
 \\ \hline
   \begin{tabular}{@{}c@{}}  Steady \\ (A)  \end{tabular}            
    & 500  
    & 1.0000  
    &
    & \begin{tabular}{@{}l@{}} $\sigma_1 = 0.0805$,  \quad $\sigma_2 =0.0597$ \\  $\beta_1 = 0.3266$, \quad  $\beta_2 =-0.1714$ \\
    $\gamma_1 = 3.0795$,  \quad $\gamma_2 =3.9828$ \end{tabular}  
    \\ \hline  
    \begin{tabular}{@{}c@{}}  Fluttering  \\ (B)       \end{tabular}        	   
    & 500 
    & 1.0000
    &
     &  
    \begin{tabular}{@{}l@{}} $\sigma_1 =  \ \ 1.0277$, \ \  \quad $\sigma_2 =0.8664$ \\  $\beta_1 = -0.1825, \ $ \quad  $\beta_2 =0.0493$ \\
    $\gamma_1 = \ \ 3.3679$,  \  \quad $\gamma_2 =3.5389$ \end{tabular} 
    \\ \hline  
      \begin{tabular}{@{}c@{}}      Chaotic  \\ (C)       	   \end{tabular}         
    & 600 
    & 0.4951 
    &
    &  \begin{tabular}{@{}l@{}} $\sigma_1 = 4.2387$, \  \quad $\sigma_2 =3.9821$ \\  $\beta_1 = -0.1963$, \quad  $\beta_2 =0.0925$ \\
    $\gamma_1 = \ \ 2.7701$,  \quad $\gamma_2 =3.1713$ \end{tabular} 
    \\ \hline  
    \begin{tabular}{@{}c@{}}   Tumbling \\ (D)         \end{tabular}         	   
    & 500 
    & 0.4033 
    &
    &   \begin{tabular}{@{}l@{}} $\sigma_1 = 5.4476$,  \quad $\sigma_2 = 4.9726$ \\  $\beta_1 = \ 0.0768$, \quad  $\beta_2 =0.1026$ \\
    $\gamma_1 = \ 1.5298$,  \quad $\gamma_2 =1.6316$ \end{tabular} 
    \\ \hline    
     \end{tabular}
    \end{center}
    \end{table}

\section{Results}
\label{sec:results}

To build a baseline from which to compare, we first tested  the steady falling 
in air.  We performed three sets of drops, each set corresponding to a different point in the parameter space using discs of various diameters. These 
three sets helped  establish limitations on the accuracy associated with the  drop mechanism, as well as the measurements taken by the camera. The scatterplots of the physical landing sites associated with one of the sets, set (A), can be seen in  the top row of Figure~\ref{fig:distribution}(A)  and the associated probability distribution in the bottom row of Figure~\ref{fig:distribution}(A). The distribution associated with the landing site is a very tightly distributed (almost Gaussian) distribution about the center drop-line, with variance listed in Table 1. The variance associated with these sets of runs can be thought of as a measure of  the uncertainty built into the experimental mechanism.

We then performed, using the same drop mechanism, repeated drops of the coin {in water} for each set of parameters corresponding to (B) fluttering, (C) chaotic and (D) tumbling. 
For each set of parameters, distinct drops of the coin result in distinct falling trajectories and landing positions. The landing positions are reported in the top row of Figures~\ref{fig:distribution}(B-D). The bottom row  shows the radial distribution of the associated landing positions. 
Particularly noteworthy for the fluttering and chaotic modes is the ``dip" in the histogram around the origin, suggesting that the origin is 
the one of the least likely places of landing in these falling modes. The probability of landing at the origin is zero for the tumbling mode.
A singular value decomposition of the landing position data was then performed. The resulting two eigen-directions are shown over the scatterplots in Figure~\ref{fig:distribution}(A-D) with $V_1$ and $V_2$ denoting the major and minor axes, respectively. The standard deviation, skewness {(third standardized moment)} , and kurtosis {(fourth standardized moment)} were calculated along each of these axes and are summarized in Table 1. {The skewness and kurtosis provide information about  the shape of the probability density function associated with the random landing site and  can be viewed as a simple measure of the extent to which these distributions deviate from Gaussian.
Roughly speaking, the deviation of the skewness from zero (perfectly symmetric pdf's) measures the extent to which the distribution tilts to one side or the other, with negative skewness indicating the `left' tail of the pdf is longer, and positive skewness indicating that the right `tail' is longer.
The interpretation of kurtosis is slightly more subtle.
It measures the `peakedness' of the distribution and the heaviness of its tail with a Gaussian distribution having a value of zero.
The values of the skewness and kurtosis in Table 1 imply that  the pdf's reported in Figure~\ref{fig:distribution} are close to symmetric; however, they are not Gaussian, given the large deviation of the kurtosis from zero.
}

\begin{figure}
\begin{center}
 	\begin{subfigure}[Fluttering trajectory I]
		{
		\includegraphics[width=0.23\textwidth]{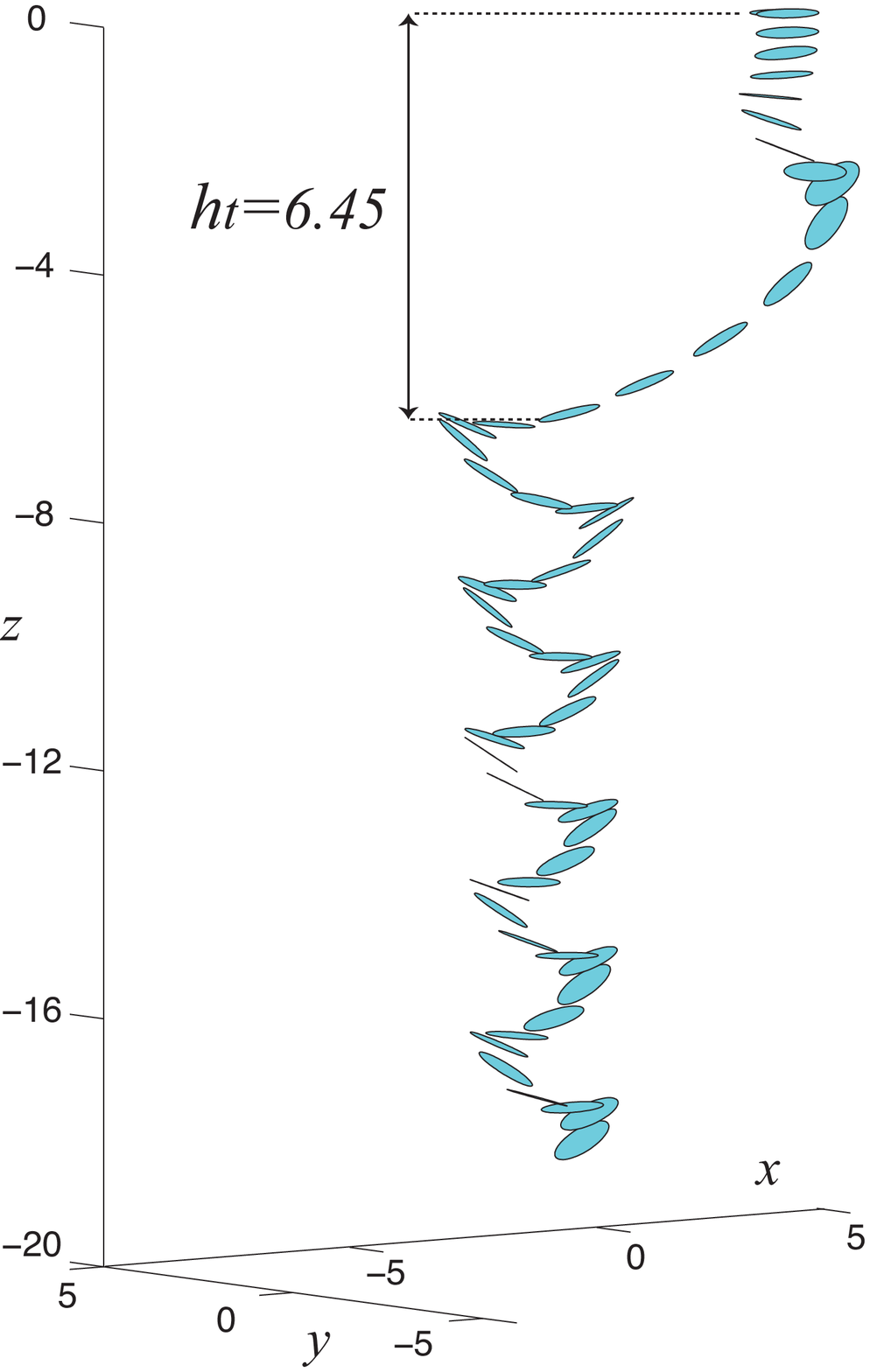}	  
		\raisebox{0.2\height}{\includegraphics[width=0.2\textwidth]{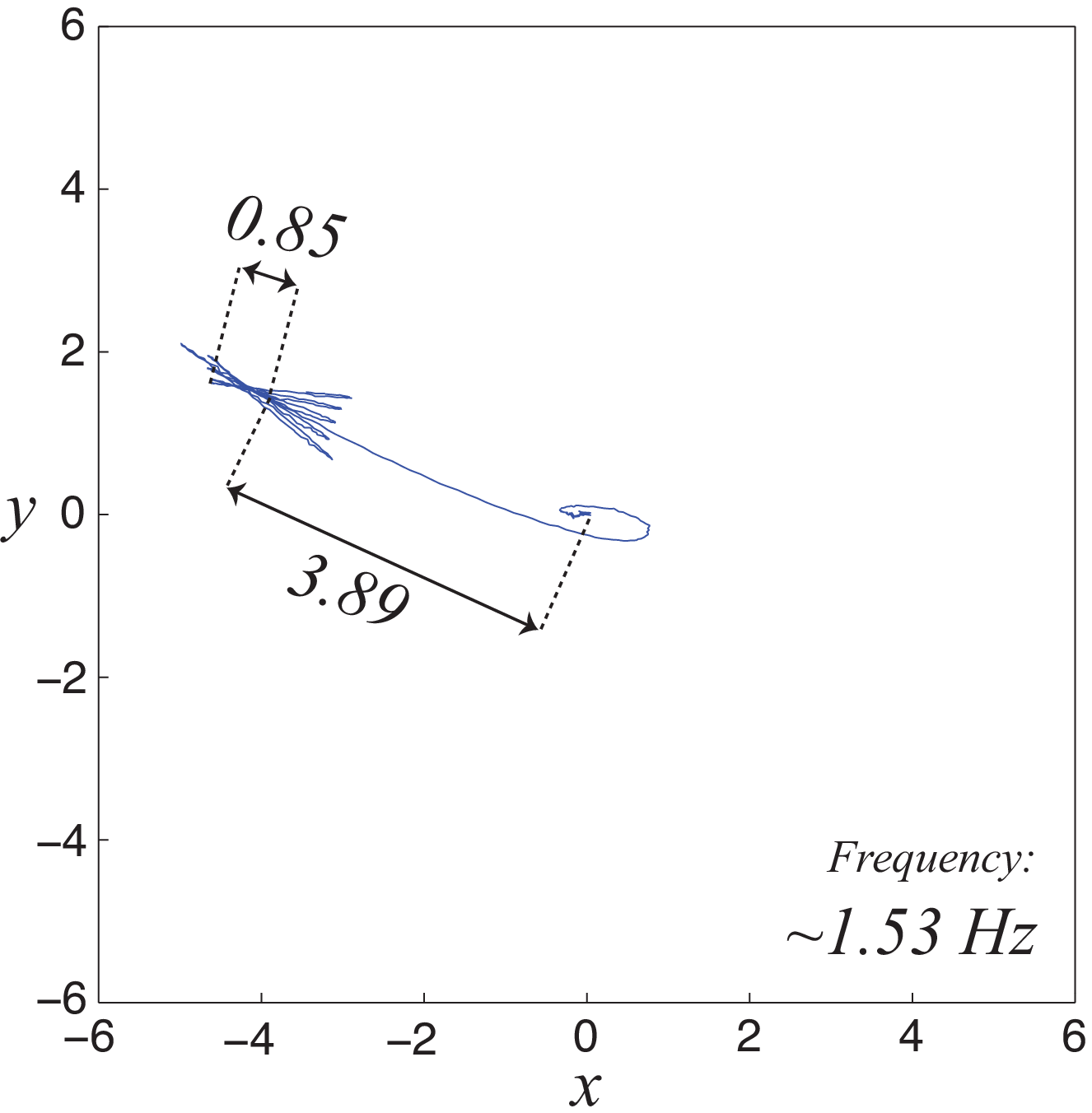}}	
		}
	\end{subfigure}\hspace{0.5in}
	 	\begin{subfigure}[Fluttering trajectory II]
		{
		\includegraphics[width=0.19\textwidth]{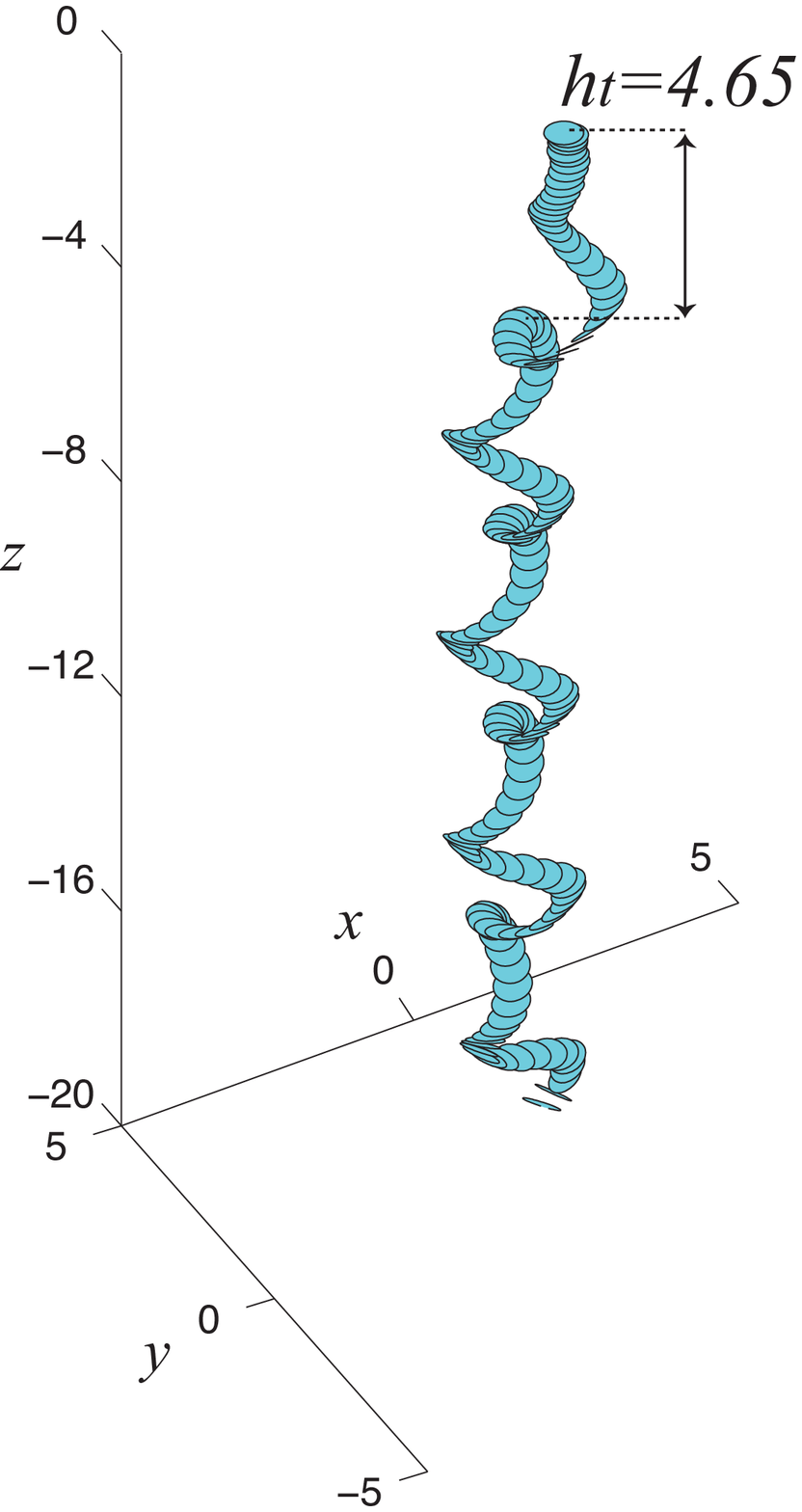} \hspace{0.1in}	  
		\raisebox{0.2\height}{\includegraphics[width=0.2\textwidth]{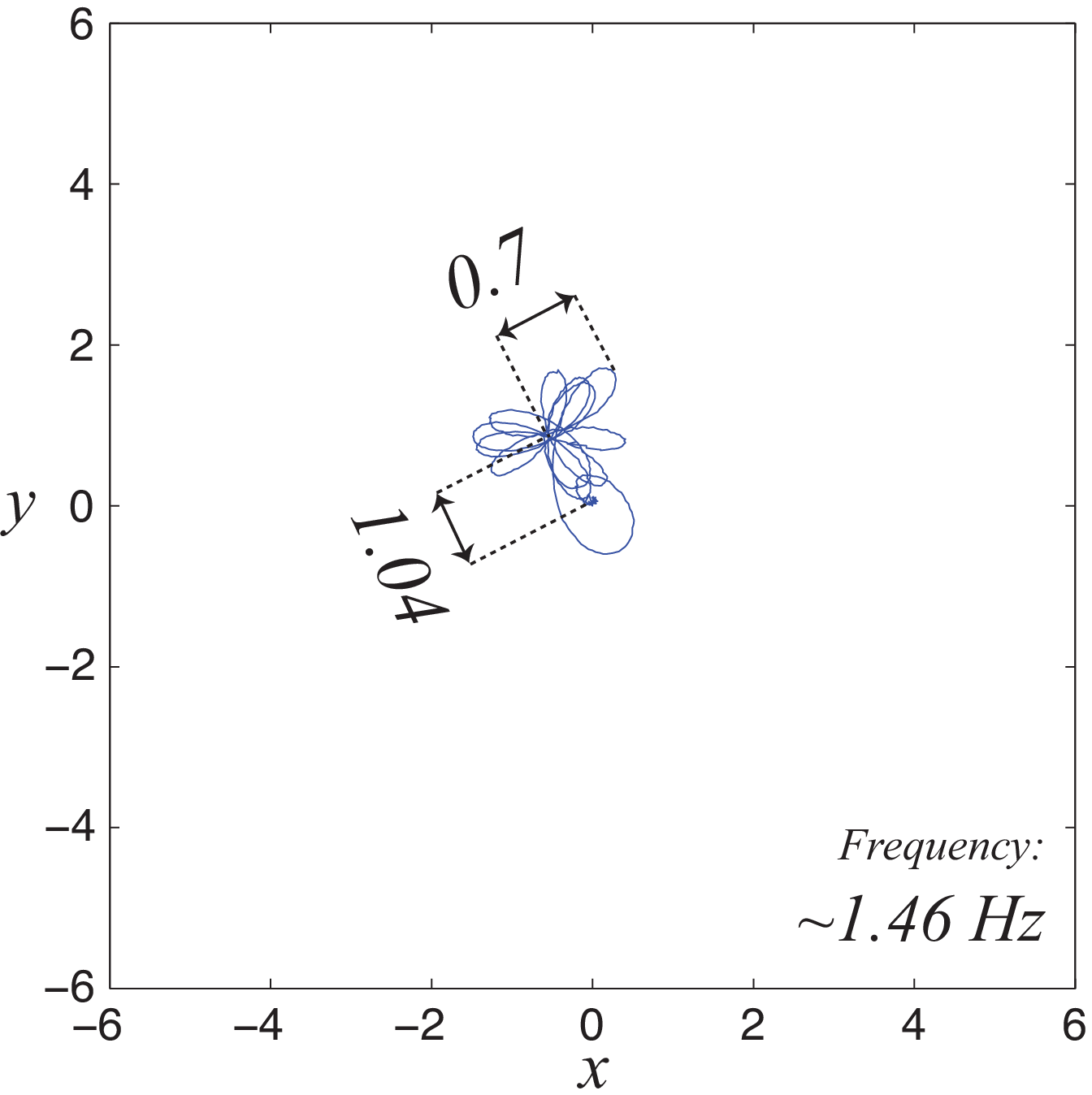}}	
		}
	\end{subfigure}
	\\[0.2in]
	\begin{subfigure}[Chaotic trajectory III]
			{
			\includegraphics[width=0.2\textwidth]{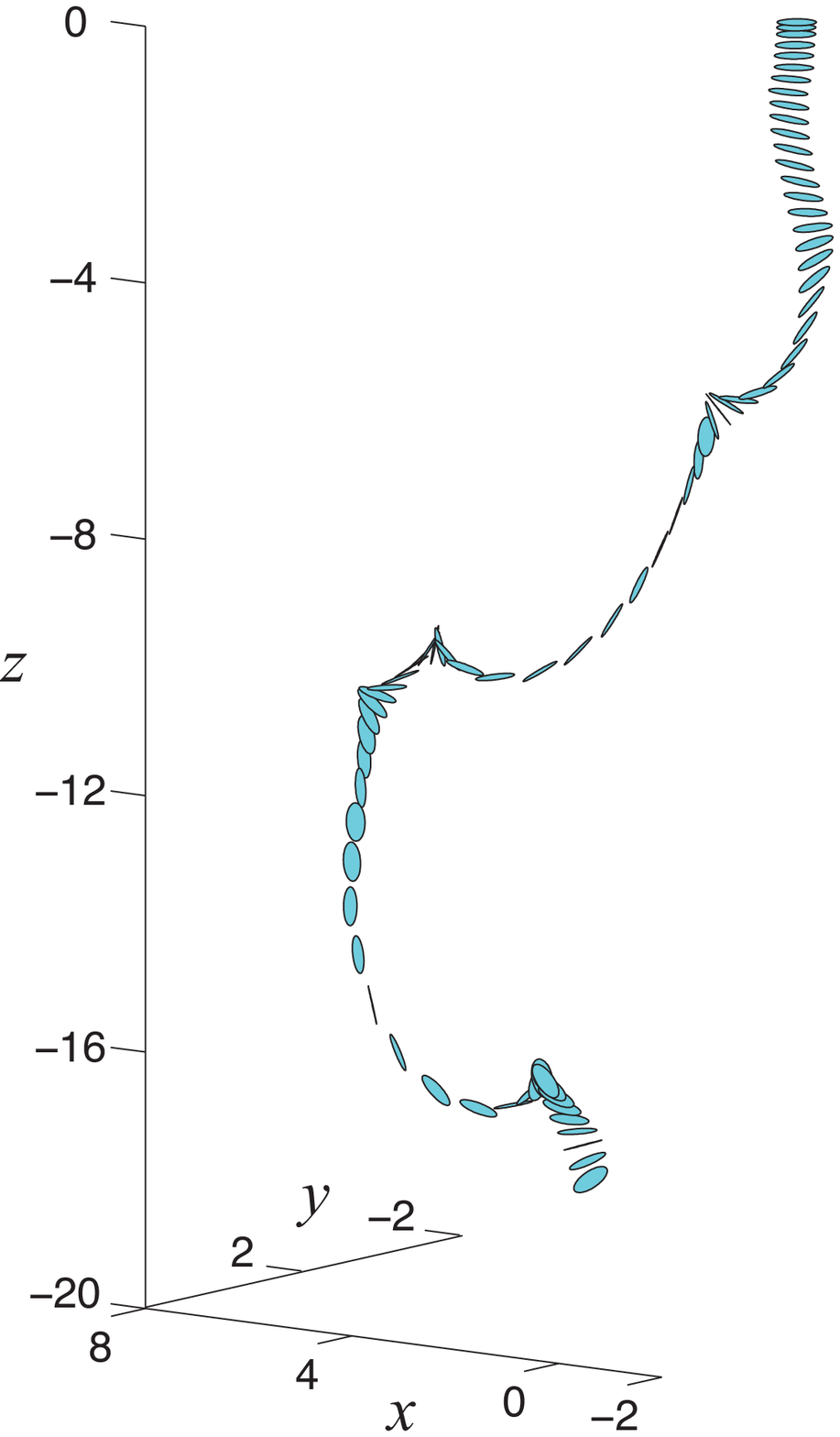}	 \hspace{0.1in}	
			\raisebox{0.2\height}{\includegraphics[width=0.2\textwidth]{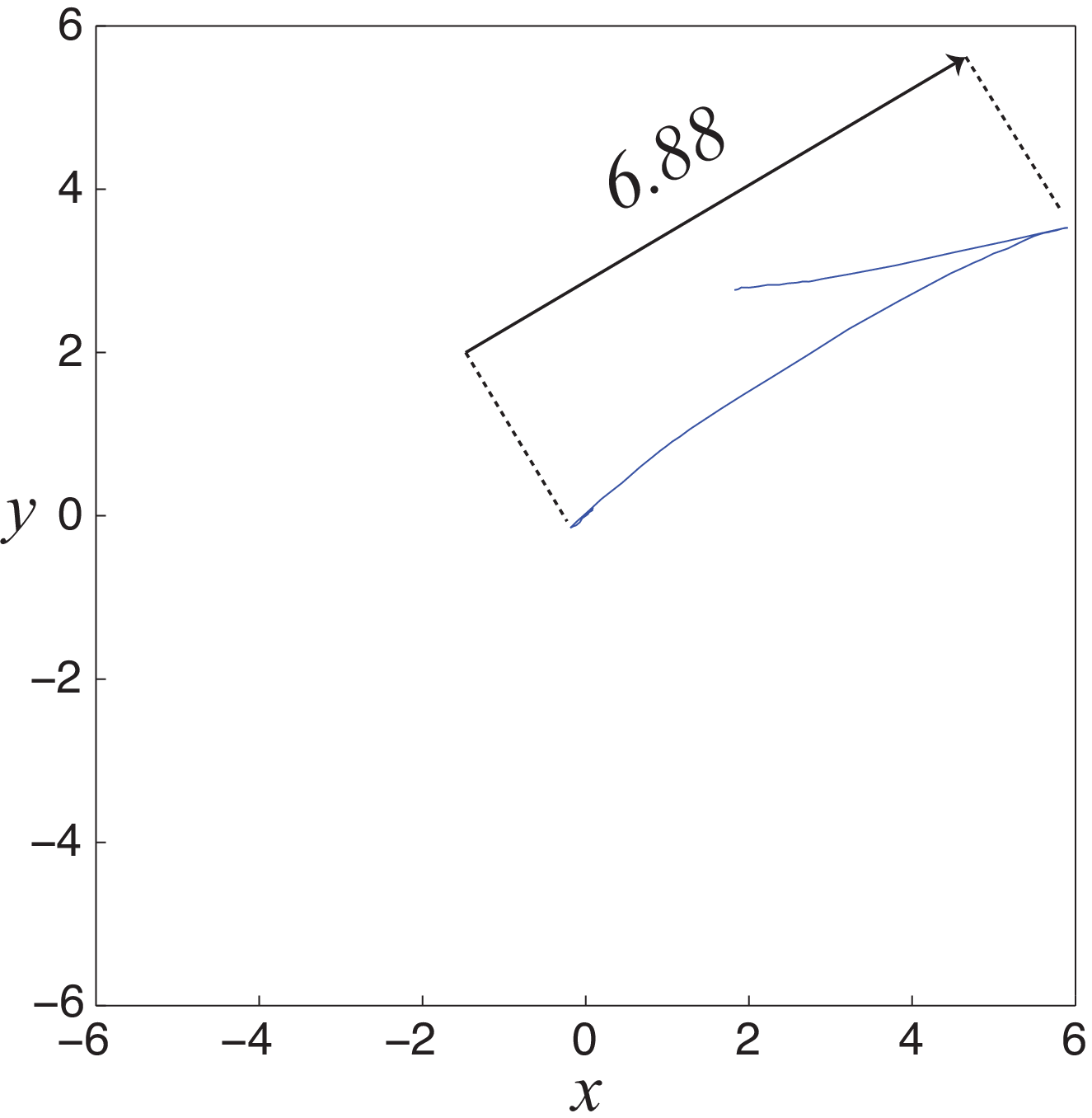}}	
			}
	\end{subfigure} \hspace{0.4in}
	\begin{subfigure}[Tumbling trajectory IV]
		{
		\includegraphics[width=0.23\textwidth]{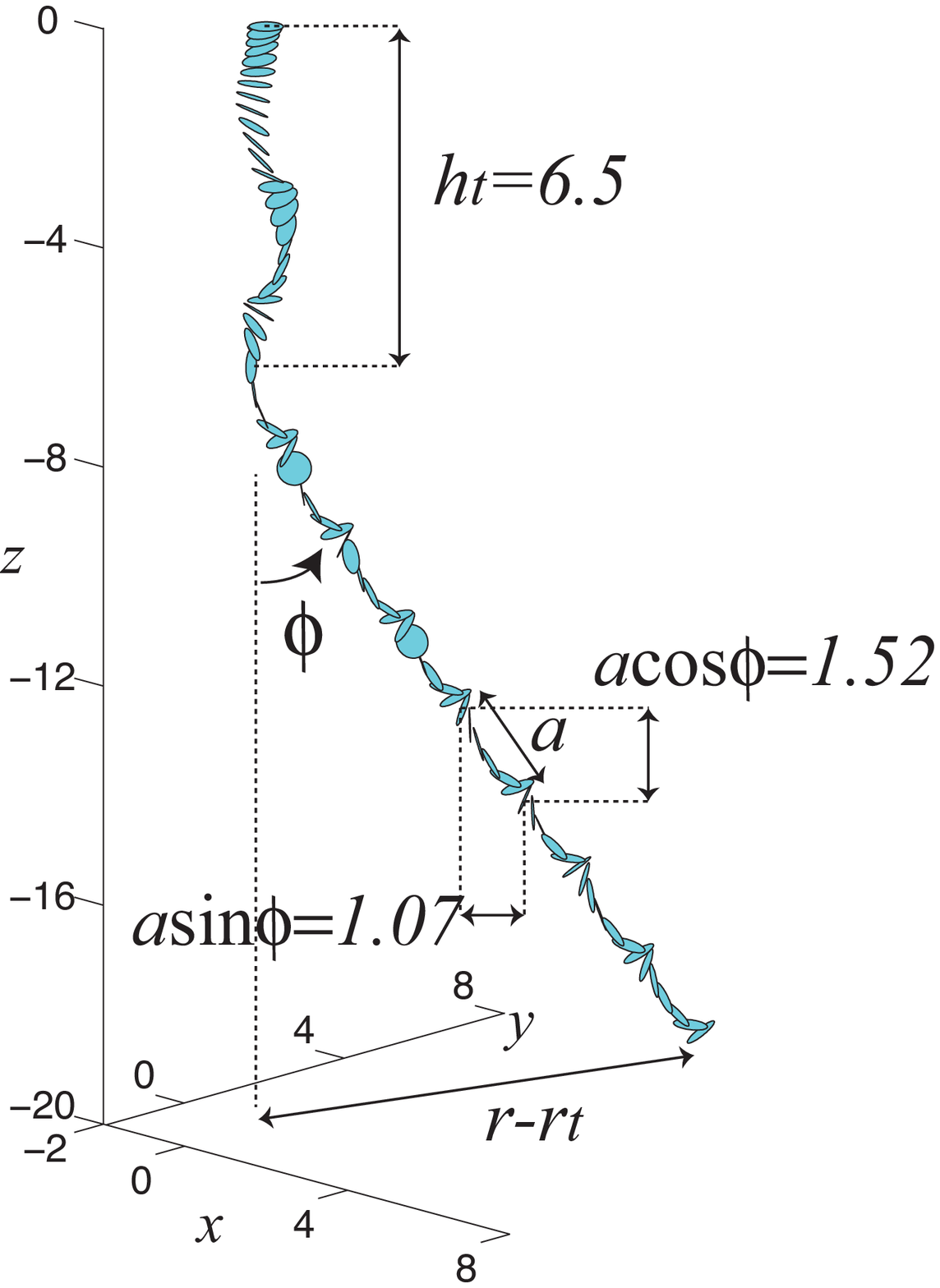}  \hspace{0.025in}		   
	         \raisebox{0.2\height}{\includegraphics[width=0.19\textwidth]{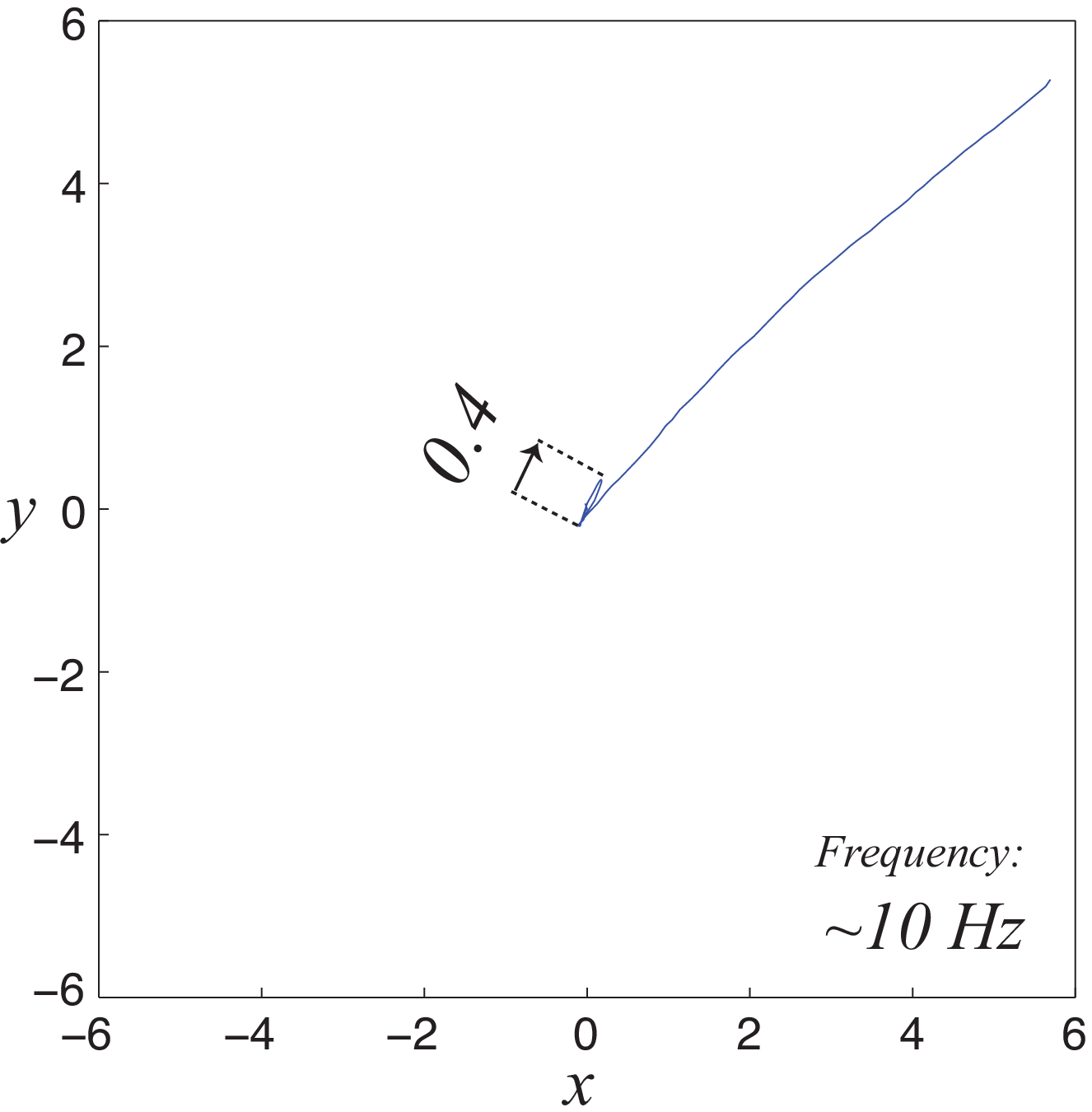}	}
		}
	\end{subfigure}
	\caption{Reconstructed 3D trajectories and $(x,y)$ planar projections. \textcolor{black}{The discs in a given figure are shown at equal intervals $\Delta t$, and thus closely spaced discs correspond to slower motion.} Drop mechanism is located at $(0,0)$. All units in inches.
	}
	\label{fig:trajectories}
\end{center}
\end{figure}

Examples of reconstructed 3D trajectories taken from these data sets are shown in Figure~\ref{fig:trajectories}.
These sample trajectories demonstrate the influence of the transient dynamics on the details of the falling trajectories and the landing sites, which in turn dictates the distributions in Figure~\ref{fig:distribution}. Figures~\ref{fig:trajectories}(a) and (b) show  two examples corresponding to two distinct falling trajectories (I) and (II) of the fluttering mode, with landing sites pointed out in Figure~\ref{fig:distribution}(B). Note
the difference in the transient dynamics and the subsequent fluttering motion. In (I), after the transience, the coin flutters from side to side in almost one plane as can be seen from the top view inset. In (II), the coin descends along an a gyrating trajectory which illustrates the three-dimensional nature of the problem. In each case, the transient displacement affects the landing position. One could think of the transient dynamics as amplifying the little uncertainty that is present in the drop mechanism (reflected in the tight distribution in Figure~\ref{fig:distribution}(A)) and thus enlarging the set of possible landing positions. Indeed, {the radius of support for which} the distribution in Figure~\ref{fig:distribution}(B) {is nonzero} is greater than the characteristic radius of oscillation  for the trajectories in Figures~\ref{fig:trajectories}(a) and (b), which is {$\approx 0.85''$ and $\approx1.04''$, respectively}. 
The transient dynamics also contributes to the ``ring" of landing positions in the tumbling case, Figure~\ref{fig:distribution}(D). However, in the chaotic case, the transient dynamics is indistinguishable. When falling in this regime, the coin seems to switch erratically between tumbling, fluttering  and steady descent. There seems to be no strong preference in the resulting landing sites, as depicted in Figure~\ref{fig:distribution}(C).

The probability of landing with the same side up as was dropped (heads) is also reported in Table 1. For the steady and fluttering modes, the coin never flips, so the coin always lands heads up. For the chaotic mode, the probability of landing heads or tails is close to 0.5. This is not surprising. Chaotic motions are characterized by lack of correlation between initial conditions and subsequent trajectory. In the case of the tumbling mode, the probability of heads or tails is based on the height of the drop which determines whether the coin flips an even or odd number of times during descent. For the parameters used in our experiment, there is a bias towards flipping an odd number of times, leading to a probability of landing heads up of about 0.4.

\textcolor{black}{We conclude this section by noting that  points  (E), (F), (G), (H) of the phase diagram in Figure~\ref{fig:fallingregimes} lead to probability distribution functions that  are similar to those reported in Figure~\ref{fig:distribution}, and therefore they are not shown here for brevity. Based on this observation, we conjecture that the distributions reported in Figure~\ref{fig:distribution} are ``generic" in the sense that they qualitatively describe the landing sites associated with each descent mode. It is important to note that the quantitative details of these distributions depend on the parameter values $(I^\ast,{\rm Re})$ in Figure~\ref{fig:fallingregimes}. However, the type of distribution (e.g., Gaussian, broadband or ringlike) seems to depend on the descent mode only and seems to be invariant for a given descent mode.}
%
%
%
%
%
%
%
%

\section{Discussion}
\label{sec:conc}

The problem of a coin dropping in water can be thought of as a canonical deterministic problem in which the delicate mechanics involved (fluid-solid interactions) generates effects that are best treated as probabilisitic. 
{The analysis of such problems goes back to the work~\cite{Poincare}  on the roulette wheel. Poincar\'{e} addressed the questions of how small variations in initial conditions and the physics of collisions determine the probabilistic outcome of the roulette wheel. Following Poincar\'{e}'s work, Hopf presented a framework for formally analyzing the role of the laws of physics in determining the flow of initial probability distribution functions to their final states; see, e.g.,~\cite{Hopf} and references therein. More recently, the coin toss problem (neglecting all aerodynamic effects) has received a great deal of attention as an example of a deterministic system with few degrees of freedom that results in random outcomes (heads or tails );
see, e.g.,~\cite{Keller1986,DiHoMo2007,YoMa2011}. These works show that the probability of heads tends to 1/2 (fair coin) only in the limit of large initial linear and angular velocities for a coin that  is spun about an axis through its plane (end-over-end spinning).  In all other situations, the outcome is biased by the coin's initial orientation.
}

In this study, our goal was to characterize the probability density function associated with the landing site for a disc falling in water, as well as the probability of landing heads, from zero and close to zero initial conditions.
{We showed that when falling in air in steady descent, the probability distribution function of the landing site is almost Gaussian, see Figure~\ref{fig:distribution}(A). We interpreted this distribution as a measure of the uncertainty that is inherent in the drop mechanism. In the absence of such uncertainty, a steadily descending disc would always land at the point right below the point from which it was dropped. We quantified the uncertainty in the drop mechanism by computing the variance, skewness and kurtosis of the resulting distribution. We  found that the variances are small (of order $\sim 0.01-0.1''$) in comparison to the disc's diameter of $1''$. It is important to point out that this order of uncertainty is typical of all drops in air. In particular, we obtained uncertainties of the same order when we took the discs of points (B), (C), (D) of Figure~\ref{fig:fallingregimes} and dropped them in air. Therefore, we view the uncertainty in the drop mechanism reported in Figure~\ref{fig:distribution}(A) as a measure of the uncertainty in the initial conditions for points (B), (C), (D). The probability distribution functions in Figures~\ref{fig:distribution}(B), (C) and (D) can be interpreted as the outcome of the dynamical evolution of this very tight, initial (almost Gaussian) probability distribution under the system's nonlinear dynamics.} 

The shape of the {probability distribution functions} reported in Figure~\ref{fig:distribution} depends intimately on the type of descent mode the object undergoes, and this is dependent on both the moment of inertia of the object, as well as the Reynolds number of the surrounding fluid. Clearly, our first intuition, that the distribution associated with landing sites would be generically Gaussian, with Reynolds number dependent variance, proved to be too simplisitic. {In this simplistic scenario,} the spot directly below the drop site would be the most likely location for the disc to land. In reality, the location directly below the drop site could, in fact, be one of the least likely spots for the disc to land. 
{For example, in the fluttering mode, the pdf is characterized by  a ``dip" at the origin. The presence of this dip is due to factors such as the transient dynamics but also to the nature of the fluttering motion. In this descent mode, \textcolor{black}{the non-vertical component of the disc's translational velocity reaches zero} as the disc reverses its direction of motion from one side to the other (see Figure~\ref{fig:trajectories}(a)), and thus spends more time close to the points of maximum amplitude. This, in turn, implies that the disc is more likely to land when it is close to these extreme points than elsewhere along its descent trajectory. 
The probability of landing heads is $1$ as the disc never flips in this descent mode.} 

{In the chaotic mode, our data suggest that the fluid medium acts as a ``randomization device" in which successive drops of the disc seem to be completely independent and the probability of falling heads is unbiased by the initial condition of heads up. This statement remains to be rigorously proven using appropriate mixing measures on the phase space of probability distribution functions. 
The data shown in Figure~\ref{fig:distribution} reveal that the pdf associated with the chaotic mode is characterized by a dip around the origin that extends over a larger distance $r$ than that of the fluttering mode and a smaller distance $r$ than that of the tumbling mode. This is due 
to the nature of chaotic descent in which the disc switches between fluttering and tumbling with potentially intermittent intervals of steady falling.}


\textcolor{black}{In the tumbling mode, the disc settles, after some transience, into revolving around itself while descending on a path that is inclined to the vertical at a nearly constant angle $\phi$ (see Figure~\ref{fig:trajectories}(d)). The radius $r$ of the landing point is therefore a function of the height $h$  from which the disc is dropped, the properties of the tumbling motion (the angle $\phi$ and the  distance spanned between two consecutive tumbles $a$), and the transient motion.  The probability of landing heads is a function of the same parameters, and thus it may be correlated to the landing site. The degree of correlation is affected by the randomness of the transient motion. Considering that for tumbling motions the trajectory of the disc's center is planar or nearly planar and letting $h_t$ and $r_t$ be the height and radius (projected onto the plane of motion) associated with the transient motion, one can then write (see Figure~\ref{fig:trajectories}(d))
\begin{equation*}\label{eq:correlation}
\begin{split}
r-r_t = (h-h_t) \tan(\phi),  \qquad n & = {\rm floor}\left[\dfrac{h-h_t}{a\cos(\phi)}\right] = 
{\rm floor}\left[\dfrac{r-r_t}{a\sin(\phi)}\right] .
\end{split}
\end{equation*}
Here,  $n\in \mathbb{N}$ is the number of times the disc revolves around itself. The disc lands heads up if $n$ is even ($n=2m$ where $m\in \mathbb{N}$), or equivalently,  if $2m\pi - \pi/2 < \left. \theta \right|_{\rm landing} < 2m\pi + \pi/2$.  We emphasize that $h_t$ and $r_t$ are random variables due to the uncertainty in the initial conditions, and consequently, $a$, $\phi$, $n$ are also random.
The above equations suggest that there is a correlation between the landing site and the probability of landing heads. This correlation between the landing site and the probability of landing head implies that one could, in principle, ``design" and build discs that, for a given experimental set-up (height $h$), produce a desired probability of landing heads or a desired mean drift in their landing site.}

{In summary, we have shown that the solid-fluid interaction in the problem of coins falling in water provides a natural mechanism for producing randomness in the landing sites of the disc and the probability of landing heads up. This randomness is due to the amplification of small variations in the initial conditions by the fluid medium which couples the translational and rotational motion of the disc. Indeed, in the absence of the fluid medium, the disc always lands heads up in a tight, almost Gaussian distribution of landing sites. This is a major difference between our work and the works of ~\cite{Keller1986,DiHoMo2007,YoMa2011} on the coin toss where the fluid effects are neglected and the initial conditions take on much larger ranges of values that produce the randomness in the toss outcome.}

Future extensions of this work will include a more quantitative description of the change 
in the pdf characteristics (mean, variance, shape/higher-order moments) as functions of the system's parameters ($I^\ast$ and Re).
It will also address the inverse problem of how to design objects that result in desired landing sites and desired probabilities of landing heads. \textcolor{black}{This undertaken will be relevant to a number of applications such as understanding the parameters influencing the range and accuracy of the passive flight
of gliding animals, see, e.g.,~\cite{PaMa2011} and references therein.} \\


\par\noindent

The work of LH and EK is partially supported by the NSF CAREER award CMMI 06-44925 and the grant CCF08-11480.  The work of PN is partially supported by NSF grant DMS08-04629.



\begin{thebibliography}{1}




\bibitem[Alben (2010)]{Alben2010}
Alben, S. 2010 Flexible sheets falling in an inviscid fluid. {\em Phys. Fluids} \textbf{22}, 061901.

\bibitem[Andersen et al. (2005)]{AnPeWa2005}
Andersen, A., Pesavento, U. \& Wang, Z. J.  2005 Unsteady aerodynamics of fluttering and tumbling plates.  {\em J. Fluid Mech.},  \textbf{541}, 65-90.


\bibitem[Andersen et al. (2005b)]{AnPeWa2005b}
Andersen, A., Pesavento, U. \& Wang, Z. J. 2005 Analysis of transitions between fluttering, tumbling and steady descent of falling cards. {\em J. Fluid Mech.}  \textbf{541}, 91-104.


\bibitem[Belmonte et al. (1998)]{BeEiMo1998}
Belmonte, A., Eisenberg, H. \& Moses, E.  1998 From flutter to tumble: inertial drag and Froude similarity in falling paper. {\em Phys. Rev. Lett.}, \textbf{81},  345-348.

\bibitem[Diaconis et al. (2007)]{DiHoMo2007}
Diaconis, P., Holmes,  S., \& Montgomery, R. 2007 Dynamical bias in the coin toss. {\em SIAM Rev.},  \textbf{49}, 211-235.


\bibitem[Dupleich (1941)]{Dupleich1941}
Dupleich, P.  1941 Rotation in free fall of rectangular wings of elongated shape. {\em NACA Tech. Memo.}  \textbf{1201}, 1-99.



\bibitem[Field et al. (1997)]{Field1997}
{Field, S. B. , Klaus, M., Moore, M. G. \& Nori, F.}  1997 {Chaotic dynamics of falling disks}, {\em Nature},  \textbf{388}, 252-254.


\bibitem[Hopf (1934)]{Hopf}
Hopf E. 1934 On causality, statistics and probability, {\em  J. Math. Phys.}  \textbf{13}, 51-102.


\bibitem[Howe \& Smallwood (1982)]{HoSm1982}
Howe, H. F. \&  Smallwood, J. 1982 Ecology of seed dispersal,
{\em Ann. Rev. Ecol.  Syst.}
 \textbf{3}, 201-228.

\bibitem[Jones (2003)]{jones} 
Jones, M. A. 2003 {The separated flow of an inviscid
fluid around a moving flat plate}, {\em J.\ Fluid Mech.} {\bf 496} ,
405--441.

\bibitem[Jones \& Shelley (2005)]{jonesshelley05}
Jones, M.\ A.\  and Shelley, M.\ J.\  2005
{Falling cards}, {\em J.\ Fluid Mech.} {\bf 540},
393--425.


\bibitem[Keller (1986)]{Keller1986}
Keller, J. B. 1986 The probability of heads. {\em Am. Math. Monthly}, \textbf{93}, 191-197.




\bibitem[Mahadevan (1996)]{Mahadevan1996}
Mahadevan, L.   1996 Tumbling of a falling card. {\em C.R. Acad. Sci. Ser. IIb}, \textbf{323}, 729-736.


\bibitem[Mahadevan et al. (1999)]{Mahadevan1999}
Mahadevan, L.   Ryu, W. S., \&  Samuel, A. D. T. 1999 Tumbling cards. {\em Phys. Fluids}, \textbf{11}(1), 1-3.


\bibitem[Maxwell (1854)]{Maxwell1854}
Maxwell, J. C. 1854  On a particular case of the descent of a heavy body in a resisting medium. {\em Camb. Dublin Math. J.} \textbf{9}, 145-148.

\bibitem[Michelin \& Llewellyn Smith (2009)]{michel09}
Michelin S. \&  Llewellyn Smith, S.\ G.\  2009
{An unsteady point vortex method for coupled fluid-solid problems},
{\sl Theor.\ Comput.\ Fluid Dyn.}\ {\bf 23}, 127--153.



\bibitem[Mittal et al. (2004)]{MiSeUd2004}
Mittal, R., Seshadri, V. \& Udaykumar, H. S. 2004 Flutter, tumble, and vortex induced autorotation
{\em Theoret. Comput. Fluid Dyn.} \textbf{17}, 165-170.



\bibitem[Nathan \& Muller-Landau (2000)]{NaMu2000}
Nathan, R., \&  Muller-Landau, H.C. 2000 Spatial patterns of seed dispersal, their determinants and consequences for recruitment. {\em Trends in Ecology \& Evolution}, \textbf{15}(7), 278-285.





\bibitem[Paoletti \& Mahadevan (2011)]{PaMa2011}
Paoletti, P. \& Mahadevan, L. 2011 Planar controlled gliding, tumbling and descent. {\em J. Fluid Mech.}, \textbf{689}, 489-516.



\bibitem[Pesavento \& Wang (2004)]{PeWa2004}
Pesavento, U. \& Wang, Z. J. 2004 Falling Paper: Navier-Stokes Solutions, Model of Fluid
Forces, and Center of Mass Elevation. {\em Phys. Rev. Lett.}, \textbf{93}(14), 144501.


\bibitem[Poincar\'{e} (1912)]{Poincare}
Poincar\'{e}, H. 1912 {\em Calcul des probabilit\'es}, Gauthier-Villars, Paris.


\bibitem[Smith (1971)]{Smith1971}
Smith, E. H.  1971 Autorotating wings: an experimental investigation. {\em J. Fluid Mech.} \textbf{50}, 513-534. 


\bibitem[Tam et al. (2010)]{TaBu2010} 
Tam, D., Bush, J. W. M., Robitaille, M. \& Kudrolli, A. {2010}
Tumbling Dynamics of Passive Flexible Wings
{\em Phys. Rev. Lett.}, \textbf{104}({18}), {184504}.

\bibitem[Tanabe \& Kaneko (1994)]{TaKa1994}
Tanabe, Y. \& Kaneko, K.  1994 Behavior of a Falling Paper. {\em Phys. Rev. Lett.}, \textbf{73}(10), 1372-1375.

\bibitem[Willmarth et al. (1964)]{Willmarth1964}
Willmarth, W. W.,  Hawk, N. E., \& Harvey, R. L. 1964 Steady and unsteady motions and wakes of freely falling disks, {\em Phys. Fluids}, \textbf{7}(2), 197-208.

\bibitem[Yong \& Mahadevan (2011)]{YoMa2011}
Yong, E. H., \& Mahadevan, L. 2011 Probability, geometry and dynamics in the toss of a thick coin. {\em Am. J. Phys.}, \textbf{79}(12), 1195-1201.


\bibitem[Zheng \& Jing (2006)]{ZhJi2006}
Zheng, L., \& Jing, W. 2006 Application of predictive guidance to re-entry vehicles. {\em 1st International Symposium on Systems Control in Aerospace and Astronautics ISSCAA 2006}, 659-663.

\end{thebibliography}
\end{document}